\documentclass{PoS}

\usepackage{graphicx}
\usepackage{dcolumn}
\usepackage{bm}

\title{The charmed-strange meson spectrum from overlap fermions on
          domain wall dynamical fermion configurations}

\ShortTitle{Charmed strange mesons on DWF configurations}

\author{\speaker{S.J. Dong}, chi QCD Collaboration \\
        Dept. of Physics and Astronomy, University of Kentucky, Lexington, KY 40506, USA\\
        E-mail: \email{super124@uky.edu}}

\author{ A. Alexandru\\
        Physics Department, The George Washington University, Washington DC
20052, USA\\
        E-mail: \email{aalexan@qwu.edu}}
\author{T. Draper,  K.F. Liu,\\
        Dept. of Physics and Astronomy, University of Kentucky, Lexington, 
KY 40506, USA\\
        E-mail: \email{draper@pa.uky.edu, liu@pa.uky.edu}}
\author{A. Li\\
        Dept. of Physics, Duke University, Durham, NC 27708, USA\\
        E-mail: \email{anyi.li@duke.edu}}
\author{T. Streuer\\
        Institute of Theoretical Physics, University of Regensburg, Germeny\\
        E-mail: \email{thomas.streuer@physik.uni-regensburg.de}}
\author{J.B. Zhang\\
        Dept. Physics, Zhejiang University, Hangzhou, P. R. China\\
        E-mail: \email{jbzhang08@zju.edu.cn}}

\abstract{
The charmed-strange meson spectrum is calculated with
overlap valence fermions on 2+1 flavor domain wall dynamical configurations
for $32^3\times  64$ lattices with a spatial size of 2.7 fm.
Both the charm and strange quark propagators are calculated with the overlap fermion action.
The calculated scalar meson at 2304(22) MeV and axial-vector meson at 2546(27) MeV are
in good agreement with the experimental masses of $D_{s0}^*(2317)$ and 
$D_{s1}(2536)$. 
}

\FullConference{The XXVII International Symposium on Lattice Field Theory\\
                 July 26-31 2009\\
                 Peking University, Beijing, China}
\begin{document}

In 2003, BaBar Collaboration announced the discovery of a charmed strange 
meson $D_{s0}^{~*}(2317)$ \cite{BaBar}. CLEO also reported the
observation of this particle in the same year~\cite{CLEO}. In the following plot,
we show the masses of the charmed strange mesons from the Particle Data Group, in which the
newly discovered $D_{s0}^{~*}(2317)$ is a scalar meson and $D_{s1} (2460)$ and  $D_{s1}(2536)$ are
axial-vector mesons. \\

\vspace*{-1cm}
\begin{figure}[th]
\begin{center}
\caption{The charmed strange meson spectrum from PDG} 
\vspace*{0.5cm}
\setlength{\unitlength}{0.240900pt}
\ifx\plotpoint\undefined\newsavebox{\plotpoint}\fi
\sbox{\plotpoint}{\rule[-0.200pt]{0.400pt}{0.400pt}}%
\begin{picture}(1349,540)(0,0)
\font\gnuplot=cmr10 at 10pt
\gnuplot
\sbox{\plotpoint}{\rule[-0.200pt]{0.400pt}{0.400pt}}%
\put(181.0,75.0){\rule[-0.200pt]{4.818pt}{0.400pt}}
\put(161,75){\makebox(0,0)[r]{1600}}
\put(1308.0,75.0){\rule[-0.200pt]{4.818pt}{0.400pt}}
\put(181.0,146.0){\rule[-0.200pt]{4.818pt}{0.400pt}}
\put(161,146){\makebox(0,0)[r]{1800}}
\put(1308.0,146.0){\rule[-0.200pt]{4.818pt}{0.400pt}}
\put(181.0,217.0){\rule[-0.200pt]{4.818pt}{0.400pt}}
\put(161,217){\makebox(0,0)[r]{2000}}
\put(1308.0,217.0){\rule[-0.200pt]{4.818pt}{0.400pt}}
\put(181.0,288.0){\rule[-0.200pt]{4.818pt}{0.400pt}}
\put(161,288){\makebox(0,0)[r]{2200}}
\put(1308.0,288.0){\rule[-0.200pt]{4.818pt}{0.400pt}}
\put(181.0,358.0){\rule[-0.200pt]{4.818pt}{0.400pt}}
\put(161,358){\makebox(0,0)[r]{2400}}
\put(1308.0,358.0){\rule[-0.200pt]{4.818pt}{0.400pt}}
\put(181.0,429.0){\rule[-0.200pt]{4.818pt}{0.400pt}}
\put(161,429){\makebox(0,0)[r]{2600}}
\put(1308.0,429.0){\rule[-0.200pt]{4.818pt}{0.400pt}}
\put(181.0,500.0){\rule[-0.200pt]{4.818pt}{0.400pt}}
\put(161,500){\makebox(0,0)[r]{2800}}
\put(1308.0,500.0){\rule[-0.200pt]{4.818pt}{0.400pt}}
\put(181.0,40.0){\rule[-0.200pt]{276.312pt}{0.400pt}}
\put(1328.0,40.0){\rule[-0.200pt]{0.400pt}{110.814pt}}
\put(181.0,500.0){\rule[-0.200pt]{276.312pt}{0.400pt}}
\put(40,270){\makebox(0,0){\rotatebox{90}{Mass (MeV)}}}
\put(304,5){\makebox(0,0){\small $D_{s2}(2^+)$}}
\put(460,5){\makebox(0,0){\small $D_s(0^{-})$}}
\put(640,5){\makebox(0,0){\small $D_{s0}^*(0^+)$}}
\put(795,5){\makebox(0,0){\small $D_{s1}(1^+)$}}
\put(959,5){\makebox(0,0){\small $D_{s1}(1^+)$}}
\put(1123,5){\makebox(0,0){\small $D_s^*(1^-?)$}}
\put(304,454){\makebox(0,0){\footnotesize $D_{s2}(2573)$}}
\put(427,242){\makebox(0,0){\footnotesize $D_{s}(1969)$}}
\put(755,415){\makebox(0,0){\footnotesize $D_{s1}(2460)$}}
\put(918,443){\makebox(0,0){\footnotesize $D_{s1}(2536)$}}
\put(591,366){\makebox(0,0){\footnotesize $D_{s0}^*(2317)$}}
\put(1082,292){\makebox(0,0){\footnotesize $D_{s}^*(2112)$}}
\put(181.0,40.0){\rule[-0.200pt]{0.400pt}{110.814pt}}
\put(263,420){\usebox{\plotpoint}}
\put(263.0,420.0){\rule[-0.200pt]{19.754pt}{0.400pt}}
\put(427,206){\usebox{\plotpoint}}
\put(427.0,206.0){\rule[-0.200pt]{19.754pt}{0.400pt}}
\put(591,329){\usebox{\plotpoint}}
\put(591.0,329.0){\rule[-0.200pt]{19.754pt}{0.400pt}}
\put(755,380){\usebox{\plotpoint}}
\put(755.0,380.0){\rule[-0.200pt]{19.513pt}{0.400pt}}
\put(918,406){\usebox{\plotpoint}}
\put(918.0,406.0){\rule[-0.200pt]{19.754pt}{0.400pt}}
\put(1082,255){\usebox{\plotpoint}}
\put(1082.0,255.0){\rule[-0.200pt]{19.754pt}{0.400pt}}
\end{picture}
\end{center}
\end{figure}
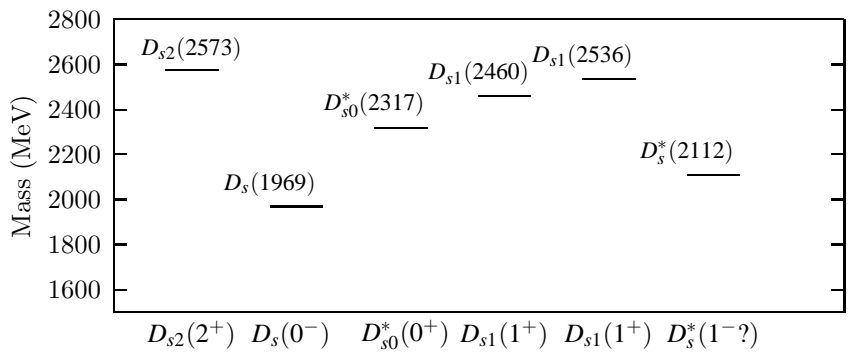

 Predictions of this charmed strange meson spectrum have been made in the 
quark model~\cite{Godfrey}. 
While it gives a good prediction for the tensor, ${}^3P_1$ axial-vector, 
pseudoscalar, and
vector mesons, its prediction of the ${}^1P_1$ axial-vector at 2.53 GeV is $\sim 70$ MeV above
the experimental $D_{s1} (2460)$. More puzzling is the prediction of the scalar meson at
2.48 GeV which is $\sim 160$ MeV above $D_{s0}^{~*}(2317)$. This discrepancy has prompted
 speculations that $D_{s0}^{~*}(2317)$ is a DK molecule~\cite{bcl03}, a four quark 
state~\cite{Cheng}, or a threshold effect~\cite{br03} instead of a $c\bar{s}$ meson.

 There are also a few lattice calculations. Lattice 
NRQCD calculation with quenched approximation gives $m(D_{s0}^*)=2.44(5)$ GeV \cite{Lewis}. 
The $n_f=2$ calculation with the heavy quark at the static limit gives 
$m(D_{s0}^*)=2.57(11)$ GeV~\cite{Gunnar}. These are also 
significantly heavier than the experimental mass of 2.317 GeV. The recent
calculation with a relativistic heavy quark (RHQ) action gives 
$\Delta m=m(D_{s0}^*)-m(D_s)
=0.1243(28)$ GeV, or $m(D_{s0}^*)=2.093(3)$ GeV~\cite{CP-PACS}, which is
significantly lower than the experimental mass of $D_{s0}^{~*}$. The first calculation with
overlap valence on 2+1 flavor domain wall fermion configurations was carried out for the 
charmed-strange mesons~\cite{amt06}. It is found that at the smallest sea mass, the $D_{s0}^*$ 
is 70(40) MeV higher than the experimental result and the hyperfine splitting is also
higher than the experimental one. This is likely due to the $O(m^2a^2)$ error with 
charm masses as heavy as $am_c = 0.72$ and 0.9 in this study.  

 Although, in principle, lattice QCD is an ideal tool to calculate the hadron
spectrum from first principles, in practice it suffers from 
systematic errors such as due
to discretization effects. These errors can be large for fermion actions
which do not have chiral symmetry at finite lattice spacing. In particular,
the $ma$ errors can be substantial for heavy quarks at the commonly used
lattice spacing $a \sim 0.1$ fm.

 The overlap fermion action obeys chiral symmetry at finite lattice
spacing and is, thus, free of $O(a)$ and $O(ma)$ errors. It is shown
that the effective quark propagator of the massive overlap fermion has the same
form as that of the continuum\cite{Liu1}; examining the dispersion relations and the hyperfine
splittings, it is found that one could use $ma \leq 0.5$ and still keep the 
$O(m^2a^2)$ errors to less than 3\% to 4\% on quenched lattices \cite{Draper}\cite{Liu2}.
In the case of the 2+1 flavor dynamical domain-wall fermion configurations with
HYP smearing, the range of $ma$ with small deviation from the expected $1/\sqrt{ma}$ behavior 
for the hyperfine splitting is extended farther than that of the quenched case. With $\rho = 1.5$ for
the overlap fermion, we find that the hyperfine splitting for heavy quarks 
fits well with the form $\frac{a}{\sqrt{ma}} + \frac{b}{\sqrt{ma}^3}$ for the range
$0.2 < ma < 0.8$ with deviation as small as  0.4\% at $ma =0.8$ and starting to grow for $ma > 0.8$. 
In the case of the $32^3 \times 64$ lattice that we work on, the charm mass corresponds
to $ma = 0.484$ which is well within the range with negligible $m^2 a^2$ error.

For comparison purposes, we first present our results on charmed strange mesons masses on a quenched
$16^{~3}\times 72$ lattice with Wilson
gauge action at $\beta = 6.3345$. With the $r_0=0.5$ fm scale, we
obtain $a=0.0560$ fm. The multi-mass overlap inverter is used to calculate propagators for 26 
quark masses for $ma$ from 0.02 to 0.85 with the charm mass at $ma = 0.431$~\cite{Sonali}\cite{Dong}.

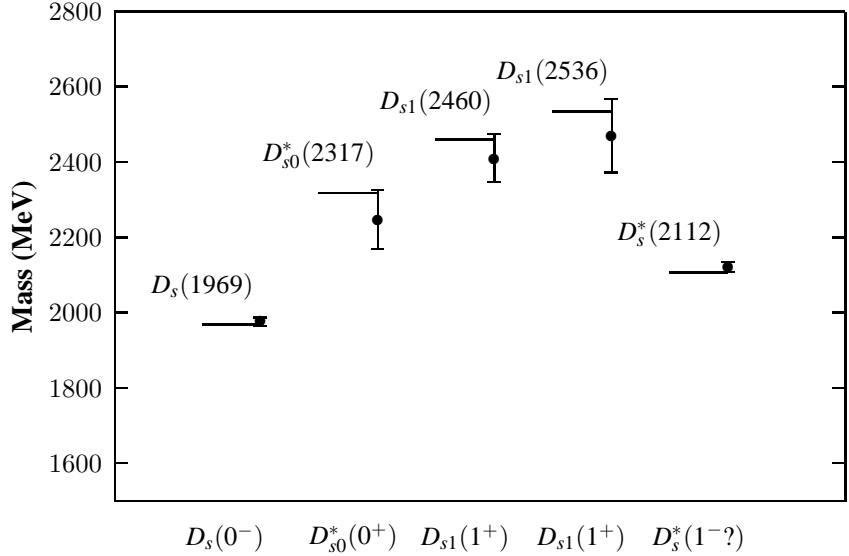
\begin{figure}[th]  
\begin{center}
\caption{$D_S$ spectrum on quenched $16^3\times 72$ lattice with overlap
fermions.}
\vspace*{0.2cm}
\setlength{\unitlength}{0.240900pt}
\ifx\plotpoint\undefined\newsavebox{\plotpoint}\fi
\sbox{\plotpoint}{\rule[-0.200pt]{0.400pt}{0.400pt}}%
\begin{picture}(1349,809)(0,0)
\font\gnuplot=cmr10 at 10pt
\gnuplot
\sbox{\plotpoint}{\rule[-0.200pt]{0.400pt}{0.400pt}}%
\put(181.0,99.0){\rule[-0.200pt]{4.818pt}{0.400pt}}
\put(161,99){\makebox(0,0)[r]{1600}}
\put(1308.0,99.0){\rule[-0.200pt]{4.818pt}{0.400pt}}
\put(181.0,217.0){\rule[-0.200pt]{4.818pt}{0.400pt}}
\put(161,217){\makebox(0,0)[r]{1800}}
\put(1308.0,217.0){\rule[-0.200pt]{4.818pt}{0.400pt}}
\put(181.0,336.0){\rule[-0.200pt]{4.818pt}{0.400pt}}
\put(161,336){\makebox(0,0)[r]{2000}}
\put(1308.0,336.0){\rule[-0.200pt]{4.818pt}{0.400pt}}
\put(181.0,454.0){\rule[-0.200pt]{4.818pt}{0.400pt}}
\put(161,454){\makebox(0,0)[r]{2200}}
\put(1308.0,454.0){\rule[-0.200pt]{4.818pt}{0.400pt}}
\put(181.0,572.0){\rule[-0.200pt]{4.818pt}{0.400pt}}
\put(161,572){\makebox(0,0)[r]{2400}}
\put(1308.0,572.0){\rule[-0.200pt]{4.818pt}{0.400pt}}
\put(181.0,691.0){\rule[-0.200pt]{4.818pt}{0.400pt}}
\put(161,691){\makebox(0,0)[r]{2600}}
\put(1308.0,691.0){\rule[-0.200pt]{4.818pt}{0.400pt}}
\put(181.0,809.0){\rule[-0.200pt]{4.818pt}{0.400pt}}
\put(161,809){\makebox(0,0)[r]{2800}}
\put(1308.0,809.0){\rule[-0.200pt]{4.818pt}{0.400pt}}
\put(181.0,40.0){\rule[-0.200pt]{276.312pt}{0.400pt}}
\put(1328.0,40.0){\rule[-0.200pt]{0.400pt}{185.252pt}}
\put(181.0,809.0){\rule[-0.200pt]{276.312pt}{0.400pt}}
\put(40,424){\makebox(0,0){\rotatebox{90}{\bf Mass (MeV)}}}
\put(355,-18){\makebox(0,0){\small\bf $D_s(0^{-})$}}
\put(557,-18){\makebox(0,0){\small\bf $D_{s0}^*(0^+)$}}
\put(732,-18){\makebox(0,0){\small\bf $D_{s1}(1^+)$}}
\put(915,-18){\makebox(0,0){\small\bf $D_{s1}(1^+)$}}
\put(1099,-18){\makebox(0,0){\small\bf $D_s^*(1^-?)$}}
\put(319,377){\makebox(0,0){\small\bf $D_{s}(1969)$}}
\put(686,667){\makebox(0,0){\small\bf $D_{s1}(2460)$}}
\put(869,714){\makebox(0,0){\small\bf $D_{s1}(2536)$}}
\put(502,584){\makebox(0,0){\small\bf $D_{s0}^*(2317)$}}
\put(1053,461){\makebox(0,0){\small\bf $D_{s}^*(2112)$}}
\put(181.0,40.0){\rule[-0.200pt]{0.400pt}{185.252pt}}
\put(319,317){\usebox{\plotpoint}}
\put(319.0,317.0){\rule[-0.200pt]{21.922pt}{0.400pt}}
\put(502,524){\usebox{\plotpoint}}
\put(502.0,524.0){\rule[-0.200pt]{22.163pt}{0.400pt}}
\put(686,608){\usebox{\plotpoint}}
\put(686.0,608.0){\rule[-0.200pt]{21.922pt}{0.400pt}}
\put(869,652){\usebox{\plotpoint}}
\put(869.0,652.0){\rule[-0.200pt]{22.163pt}{0.400pt}}
\put(1053,399){\usebox{\plotpoint}}
\put(1053.0,399.0){\rule[-0.200pt]{21.922pt}{0.400pt}}
\put(410.0,315.0){\rule[-0.200pt]{0.400pt}{3.132pt}}
\put(400.0,315.0){\rule[-0.200pt]{4.818pt}{0.400pt}}
\put(400.0,328.0){\rule[-0.200pt]{4.818pt}{0.400pt}}
\put(594.0,436.0){\rule[-0.200pt]{0.400pt}{22.404pt}}
\put(584.0,436.0){\rule[-0.200pt]{4.818pt}{0.400pt}}
\put(584.0,529.0){\rule[-0.200pt]{4.818pt}{0.400pt}}
\put(777.0,541.0){\rule[-0.200pt]{0.400pt}{18.067pt}}
\put(767.0,541.0){\rule[-0.200pt]{4.818pt}{0.400pt}}
\put(767.0,616.0){\rule[-0.200pt]{4.818pt}{0.400pt}}
\put(961.0,556.0){\rule[-0.200pt]{0.400pt}{27.703pt}}
\put(951.0,556.0){\rule[-0.200pt]{4.818pt}{0.400pt}}
\put(951.0,671.0){\rule[-0.200pt]{4.818pt}{0.400pt}}
\put(1144.0,400.0){\rule[-0.200pt]{0.400pt}{3.613pt}}
\put(1134.0,400.0){\rule[-0.200pt]{4.818pt}{0.400pt}}
\put(410,322){\circle*{18}}
\put(594,482){\circle*{18}}
\put(777,578){\circle*{18}}
\put(961,614){\circle*{18}}
\put(1144,407){\circle*{18}}
\put(1134.0,415.0){\rule[-0.200pt]{4.818pt}{0.400pt}}
\end{picture} \label{quench}
\end{center}
\end{figure}

The quenched results of the scalar and axial $D_s$, as shown in Fig.~\ref{quench}, are consistent with 
the experimental masses within error bars; albeit with relatively large errors of $\sim 80$ MeV. 
In order to better understand the $D_S$ mesons on the full QCD lattice, it is desirable to use chiral 
lattice fermions
which have smaller $O(m^2a^2)$ errors. We use valence overlap fermions on 2+1 flavor DWF dynamical 
fermion configurations with HYP smearing. This is a mixed action approach. Since the valence is a chiral
fermion, there is only one extra low-energy constant $\Delta_{mix}$ in the mixed valence-sea pion masses
which needs to be determined in the mixed action partially quenched chiral perturbation theory~\cite{JW}.
The octet pseudoscalar meson masses are thus related to the quark masses as follows:
 
\begin{eqnarray}
m^2_{v_1,v_2} &=& B(m_{v_1}+m_{v_2}),\\
m^2_{vs} &=& B(m_v+m_s)+a^2\Delta_{mix} \\
m^2_{s_1,s_2} &=& B(m_{s_1} + m_{s_2}) + a^2\Delta_{sea} \\
\Delta_{sea} &\sim & m_{res}
\end{eqnarray}

We employ 50 $N_f=2+1$ domain wall dynamical configurations from
RBC and UKQCD collaborations \cite{RBC}. The lattice has a size
$32^3\times 64$\, $L_s=16$,
with a heavier sea quark $m_h a=0.03$, close to the strange,
and light sea quarks $m_la=0.006$. The
lattice scale is $a^{-1}=2.42$GeV from $r_0 =0.47$ fm~\cite{RBC}.

\begin{figure}[th]
\begin{center}  
\caption{The equal quark mass meson spectrum on
$32^3\times 64$ lattice with $m_{sea}a=0.006$.}
\input{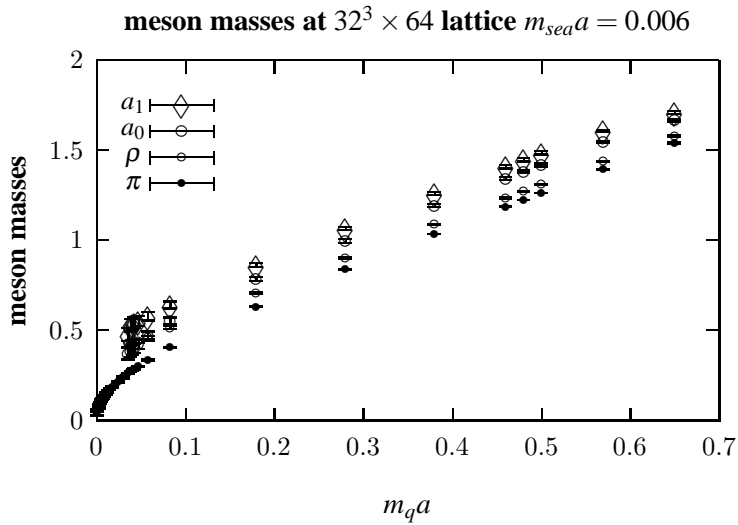}   \label{1}
\end{center}
\end{figure}

 For meson correlators, we use standard local interpolating fields.
To determine the charm quark mass $m_c a$ we use $m(J/\psi)=3097$ MeV
as the input. It gives $m_ca = 0.484$. Fig.~\ref{1} shows the meson spectrum as
a function of the quark mass. 

\begin{table}[h]
\begin{center}
\caption{\bf Charmonium meson masses }  \label{Table1}
\vspace{0.1truein}
\begin{tabular}{|c|c|c|c|c|}
\hline
Lattice & $J/\psi$(MeV) & $1^{+}$(MeV) & $0^{-}$(MeV) & $0^{+}$(MeV) \\
\hline
$32^3\times 64$ $m_la=0.006$&3097&3510(11)&2983(4)&3402(8) \\
\hline
$16^3\times 72$ quenched& 3097 &3390(50)&3017(4)&3360(50) \\
\hline
Experiment               &3097&3511    &2980   &3410 \\
\hline 
\end{tabular}  
\end{center}
\end{table}

In view of the fact that the dynamical fermion calculation is based on a much larger lattice
($32^3 \times 64$ with a spatial size of 2.7 fm) than that of the quenched lattice ($16^3 \times 72$
with a spatial size of 0.9 fm), we expect the present results to have better statistics than the 
quenched results for both the charmonium~\cite{Sonali} and the charmed strange mesons\cite{Dong} 
spectra. Indeed as we see in Table~\ref{Table1}, the $32^3 \times 64$ dynamical lattice results are 
much better than
those of the quenched results not only in statistics, but also in agreement with experiments. Even
the hyperfine splitting agrees with experiment within one sigma. 


To calculate the $D_s$ spectrum, we use the same overlap fermion
action for both light quark ($m_qa\sim 0.04$) and heavy quark
($m_qa\sim 0.484$). These heavy-light quark propagators are used to 
construct the
charmed-strange meson correlators. 
 The meson correlators of the heavy-light quarks are very well behaved
as Fig.4 shows, for example.

\begin{figure}[th]
\begin{center}
\caption{Unequal quark mass meson correlator on $32^3\times 64$ lattice}
\input{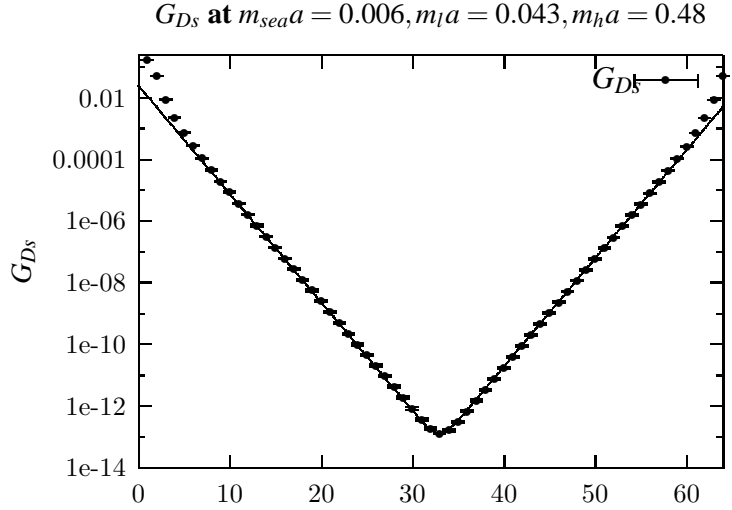}
\end{center}
\end{figure}

 To determine the charmed strange meson mass we follow these 
steps:
\begin{enumerate}
\item We use standard non-linear fitting of the pseudoscalar meson correlators for different 
$m_s$ and $m_c$. The $D_S$ mass results
are given in Table~\ref{Table2}. For each fixed $m_s a$, the meson mass is a function
of $m_c a$, as shown in Fig.5 (left).
\item Then for each $m_s a$, we interpolate data to 
$m_c a=0.484$ which was obtained previously from the $m(J/\psi)$ input.
Then the data are a function of $m_s a$ only, as shown in Fig. 5 (right).
\item By inputting 
$m_{D_s}=1969$ MeV, which corresponds to $m_{D_s} a=0.8136$, we obtain the strange quark mass
$m_s a=0.0426$.
\item The charmed strange meson masses with other quantum numbers are interpolated to $m_ca=0.484$
and $m_s a=0.0426$ to obtain the $D_s$ spectrum. They are compared with experimental data in Table 3 and
Fig. 6.
\end{enumerate}

\begin{table}[h]  
\begin{center}
\caption{\bf The light-heavy quark $D_S$ meson mass matrix} \label{Table2}
\vspace{0.1truein}
\begin{tabular}{|c|c|c|c|}  
\hline
$m_qa$ & 0.46 & 0.48 & 0.50 \\
\hline
0.039  &0.7873(28)&0.8047(28)&0.8282(29) \\
\hline
0.041  &0.7893(27)&0.8067(28)&0.8294(28) \\
\hline
0.043  &0.7912(27)&0.8086(27)&0.8302(28) \\
\hline
0.047  &0.7917(26)&0.8145(28)&0.8334(27) \\
\hline
\end{tabular}
\end{center}
\end{table}

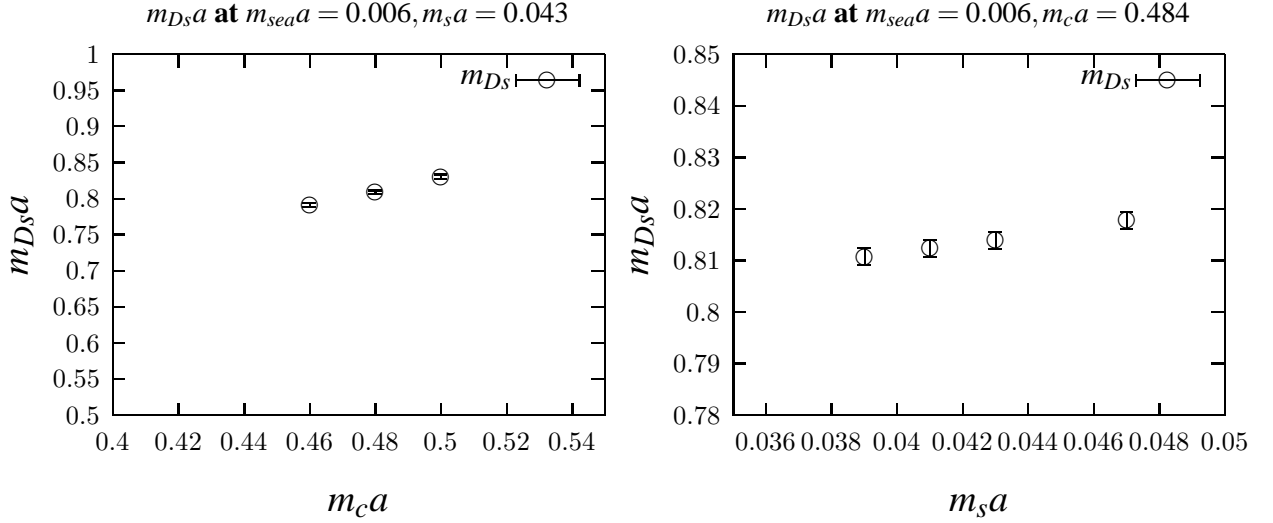
\begin{figure}[th]
\begin{center}
\caption{\bf The $D_S$ masses for several $m_c a$ with fixed $m_s a$,
and for several $m_s a$ after fixing $m_c a$ to $0.484$}
\[
\setlength{\unitlength}{0.240900pt}
\ifx\plotpoint\undefined\newsavebox{\plotpoint}\fi
\sbox{\plotpoint}{\rule[-0.200pt]{0.400pt}{0.400pt}}%
\begin{picture}(974,854)(0,0)
\font\gnuplot=cmr10 at 10pt
\gnuplot
\sbox{\plotpoint}{\rule[-0.200pt]{0.400pt}{0.400pt}}%
\put(181.0,164.0){\rule[-0.200pt]{4.818pt}{0.400pt}}
\put(161,164){\makebox(0,0)[r]{0.5}}
\put(933.0,164.0){\rule[-0.200pt]{4.818pt}{0.400pt}}
\put(181.0,221.0){\rule[-0.200pt]{4.818pt}{0.400pt}}
\put(161,221){\makebox(0,0)[r]{0.55}}
\put(933.0,221.0){\rule[-0.200pt]{4.818pt}{0.400pt}}
\put(181.0,277.0){\rule[-0.200pt]{4.818pt}{0.400pt}}
\put(161,277){\makebox(0,0)[r]{0.6}}
\put(933.0,277.0){\rule[-0.200pt]{4.818pt}{0.400pt}}
\put(181.0,334.0){\rule[-0.200pt]{4.818pt}{0.400pt}}
\put(161,334){\makebox(0,0)[r]{0.65}}
\put(933.0,334.0){\rule[-0.200pt]{4.818pt}{0.400pt}}
\put(181.0,391.0){\rule[-0.200pt]{4.818pt}{0.400pt}}
\put(161,391){\makebox(0,0)[r]{0.7}}
\put(933.0,391.0){\rule[-0.200pt]{4.818pt}{0.400pt}}
\put(181.0,448.0){\rule[-0.200pt]{4.818pt}{0.400pt}}
\put(161,448){\makebox(0,0)[r]{0.75}}
\put(933.0,448.0){\rule[-0.200pt]{4.818pt}{0.400pt}}
\put(181.0,504.0){\rule[-0.200pt]{4.818pt}{0.400pt}}
\put(161,504){\makebox(0,0)[r]{0.8}}
\put(933.0,504.0){\rule[-0.200pt]{4.818pt}{0.400pt}}
\put(181.0,561.0){\rule[-0.200pt]{4.818pt}{0.400pt}}
\put(161,561){\makebox(0,0)[r]{0.85}}
\put(933.0,561.0){\rule[-0.200pt]{4.818pt}{0.400pt}}
\put(181.0,618.0){\rule[-0.200pt]{4.818pt}{0.400pt}}
\put(161,618){\makebox(0,0)[r]{0.9}}
\put(933.0,618.0){\rule[-0.200pt]{4.818pt}{0.400pt}}
\put(181.0,674.0){\rule[-0.200pt]{4.818pt}{0.400pt}}
\put(161,674){\makebox(0,0)[r]{0.95}}
\put(933.0,674.0){\rule[-0.200pt]{4.818pt}{0.400pt}}
\put(181.0,731.0){\rule[-0.200pt]{4.818pt}{0.400pt}}
\put(161,731){\makebox(0,0)[r]{1}}
\put(933.0,731.0){\rule[-0.200pt]{4.818pt}{0.400pt}}
\put(181.0,164.0){\rule[-0.200pt]{0.400pt}{4.818pt}}
\put(181,123){\makebox(0,0){0.4}}
\put(181.0,711.0){\rule[-0.200pt]{0.400pt}{4.818pt}}
\put(284.0,164.0){\rule[-0.200pt]{0.400pt}{4.818pt}}
\put(284,123){\makebox(0,0){0.42}}
\put(284.0,711.0){\rule[-0.200pt]{0.400pt}{4.818pt}}
\put(387.0,164.0){\rule[-0.200pt]{0.400pt}{4.818pt}}
\put(387,123){\makebox(0,0){0.44}}
\put(387.0,711.0){\rule[-0.200pt]{0.400pt}{4.818pt}}
\put(490.0,164.0){\rule[-0.200pt]{0.400pt}{4.818pt}}
\put(490,123){\makebox(0,0){0.46}}
\put(490.0,711.0){\rule[-0.200pt]{0.400pt}{4.818pt}}
\put(593.0,164.0){\rule[-0.200pt]{0.400pt}{4.818pt}}
\put(593,123){\makebox(0,0){0.48}}
\put(593.0,711.0){\rule[-0.200pt]{0.400pt}{4.818pt}}
\put(696.0,164.0){\rule[-0.200pt]{0.400pt}{4.818pt}}
\put(696,123){\makebox(0,0){0.5}}
\put(696.0,711.0){\rule[-0.200pt]{0.400pt}{4.818pt}}
\put(799.0,164.0){\rule[-0.200pt]{0.400pt}{4.818pt}}
\put(799,123){\makebox(0,0){0.52}}
\put(799.0,711.0){\rule[-0.200pt]{0.400pt}{4.818pt}}
\put(902.0,164.0){\rule[-0.200pt]{0.400pt}{4.818pt}}
\put(902,123){\makebox(0,0){0.54}}
\put(902.0,711.0){\rule[-0.200pt]{0.400pt}{4.818pt}}
\put(181.0,164.0){\rule[-0.200pt]{185.975pt}{0.400pt}}
\put(953.0,164.0){\rule[-0.200pt]{0.400pt}{136.590pt}}
\put(181.0,731.0){\rule[-0.200pt]{185.975pt}{0.400pt}}
\put(40,447){\makebox(0,0){\rotatebox{90}{\Large\bf $m_{Ds} a$}}}
\put(567,21){\makebox(0,0){\Large\bf $m_c a$}}
\put(567,793){\makebox(0,0){\bf $m_{Ds} a$ at $m_{sea}a=0.006,m_sa=0.043$}}
\put(181.0,164.0){\rule[-0.200pt]{0.400pt}{136.590pt}}
\put(793,691){\makebox(0,0){\large\bf $m_{Ds}~~~$ }}
\put(813.0,691.0){\rule[-0.200pt]{24.090pt}{0.400pt}}
\put(813.0,681.0){\rule[-0.200pt]{0.400pt}{4.818pt}}
\put(913.0,681.0){\rule[-0.200pt]{0.400pt}{4.818pt}}
\put(490.0,491.0){\rule[-0.200pt]{0.400pt}{1.445pt}}
\put(480.0,491.0){\rule[-0.200pt]{4.818pt}{0.400pt}}
\put(480.0,497.0){\rule[-0.200pt]{4.818pt}{0.400pt}}
\put(593.0,511.0){\rule[-0.200pt]{0.400pt}{1.445pt}}
\put(583.0,511.0){\rule[-0.200pt]{4.818pt}{0.400pt}}
\put(583.0,517.0){\rule[-0.200pt]{4.818pt}{0.400pt}}
\put(696.0,535.0){\rule[-0.200pt]{0.400pt}{1.686pt}}
\put(686.0,535.0){\rule[-0.200pt]{4.818pt}{0.400pt}}
\put(490,494){\circle{24}}
\put(593,514){\circle{24}}
\put(696,538){\circle{24}}
\put(863,691){\circle{24}}
\put(686.0,542.0){\rule[-0.200pt]{4.818pt}{0.400pt}}
\end{picture}
\setlength{\unitlength}{0.240900pt}
\ifx\plotpoint\undefined\newsavebox{\plotpoint}\fi
\sbox{\plotpoint}{\rule[-0.200pt]{0.400pt}{0.400pt}}%
\begin{picture}(974,854)(0,0)
\font\gnuplot=cmr10 at 10pt
\gnuplot
\sbox{\plotpoint}{\rule[-0.200pt]{0.400pt}{0.400pt}}%
\put(181.0,164.0){\rule[-0.200pt]{4.818pt}{0.400pt}}
\put(161,164){\makebox(0,0)[r]{0.78}}
\put(933.0,164.0){\rule[-0.200pt]{4.818pt}{0.400pt}}
\put(181.0,245.0){\rule[-0.200pt]{4.818pt}{0.400pt}}
\put(161,245){\makebox(0,0)[r]{0.79}}
\put(933.0,245.0){\rule[-0.200pt]{4.818pt}{0.400pt}}
\put(181.0,326.0){\rule[-0.200pt]{4.818pt}{0.400pt}}
\put(161,326){\makebox(0,0)[r]{0.8}}
\put(933.0,326.0){\rule[-0.200pt]{4.818pt}{0.400pt}}
\put(181.0,407.0){\rule[-0.200pt]{4.818pt}{0.400pt}}
\put(161,407){\makebox(0,0)[r]{0.81}}
\put(933.0,407.0){\rule[-0.200pt]{4.818pt}{0.400pt}}
\put(181.0,488.0){\rule[-0.200pt]{4.818pt}{0.400pt}}
\put(161,488){\makebox(0,0)[r]{0.82}}
\put(933.0,488.0){\rule[-0.200pt]{4.818pt}{0.400pt}}
\put(181.0,569.0){\rule[-0.200pt]{4.818pt}{0.400pt}}
\put(161,569){\makebox(0,0)[r]{0.83}}
\put(933.0,569.0){\rule[-0.200pt]{4.818pt}{0.400pt}}
\put(181.0,650.0){\rule[-0.200pt]{4.818pt}{0.400pt}}
\put(161,650){\makebox(0,0)[r]{0.84}}
\put(933.0,650.0){\rule[-0.200pt]{4.818pt}{0.400pt}}
\put(181.0,731.0){\rule[-0.200pt]{4.818pt}{0.400pt}}
\put(161,731){\makebox(0,0)[r]{0.85}}
\put(933.0,731.0){\rule[-0.200pt]{4.818pt}{0.400pt}}
\put(232.0,164.0){\rule[-0.200pt]{0.400pt}{4.818pt}}
\put(232,123){\makebox(0,0){0.036}}
\put(232.0,711.0){\rule[-0.200pt]{0.400pt}{4.818pt}}
\put(335.0,164.0){\rule[-0.200pt]{0.400pt}{4.818pt}}
\put(335,123){\makebox(0,0){0.038}}
\put(335.0,711.0){\rule[-0.200pt]{0.400pt}{4.818pt}}
\put(438.0,164.0){\rule[-0.200pt]{0.400pt}{4.818pt}}
\put(438,123){\makebox(0,0){0.04}}
\put(438.0,711.0){\rule[-0.200pt]{0.400pt}{4.818pt}}
\put(541.0,164.0){\rule[-0.200pt]{0.400pt}{4.818pt}}
\put(541,123){\makebox(0,0){0.042}}
\put(541.0,711.0){\rule[-0.200pt]{0.400pt}{4.818pt}}
\put(644.0,164.0){\rule[-0.200pt]{0.400pt}{4.818pt}}
\put(644,123){\makebox(0,0){0.044}}
\put(644.0,711.0){\rule[-0.200pt]{0.400pt}{4.818pt}}
\put(747.0,164.0){\rule[-0.200pt]{0.400pt}{4.818pt}}
\put(747,123){\makebox(0,0){0.046}}
\put(747.0,711.0){\rule[-0.200pt]{0.400pt}{4.818pt}}
\put(850.0,164.0){\rule[-0.200pt]{0.400pt}{4.818pt}}
\put(850,123){\makebox(0,0){0.048}}
\put(850.0,711.0){\rule[-0.200pt]{0.400pt}{4.818pt}}
\put(953.0,164.0){\rule[-0.200pt]{0.400pt}{4.818pt}}
\put(953,123){\makebox(0,0){0.05}}
\put(953.0,711.0){\rule[-0.200pt]{0.400pt}{4.818pt}}
\put(181.0,164.0){\rule[-0.200pt]{185.975pt}{0.400pt}}
\put(953.0,164.0){\rule[-0.200pt]{0.400pt}{136.590pt}}
\put(181.0,731.0){\rule[-0.200pt]{185.975pt}{0.400pt}}
\put(40,447){\makebox(0,0){\rotatebox{90}{\Large\bf $m_{Ds} a$}}}
\put(567,21){\makebox(0,0){\Large\bf $m_s a$}}
\put(567,793){\makebox(0,0){\bf $m_{Ds} a$ at $m_{sea}a=0.006,m_ca=0.484$}}
\put(181.0,164.0){\rule[-0.200pt]{0.400pt}{136.590pt}}
\put(793,691){\makebox(0,0){\large\bf $m_{Ds}~~~$ }}
\put(813.0,691.0){\rule[-0.200pt]{24.090pt}{0.400pt}}
\put(813.0,681.0){\rule[-0.200pt]{0.400pt}{4.818pt}}
\put(913.0,681.0){\rule[-0.200pt]{0.400pt}{4.818pt}}
\put(387.0,400.0){\rule[-0.200pt]{0.400pt}{6.504pt}}
\put(377.0,400.0){\rule[-0.200pt]{4.818pt}{0.400pt}}
\put(377.0,427.0){\rule[-0.200pt]{4.818pt}{0.400pt}}
\put(490.0,413.0){\rule[-0.200pt]{0.400pt}{6.263pt}}
\put(480.0,413.0){\rule[-0.200pt]{4.818pt}{0.400pt}}
\put(480.0,439.0){\rule[-0.200pt]{4.818pt}{0.400pt}}
\put(593.0,425.0){\rule[-0.200pt]{0.400pt}{6.504pt}}
\put(583.0,425.0){\rule[-0.200pt]{4.818pt}{0.400pt}}
\put(583.0,452.0){\rule[-0.200pt]{4.818pt}{0.400pt}}
\put(799.0,457.0){\rule[-0.200pt]{0.400pt}{6.263pt}}
\put(789.0,457.0){\rule[-0.200pt]{4.818pt}{0.400pt}}
\put(387,413){\circle{24}}
\put(490,426){\circle{24}}
\put(593,439){\circle{24}}
\put(799,470){\circle{24}}
\put(863,691){\circle{24}}
\put(789.0,483.0){\rule[-0.200pt]{4.818pt}{0.400pt}}
\end{picture}
\]
\end{center}
\end{figure}

\begin{table}[h]
\begin{center}
\caption{
 Charmed strange meson masses }
\vspace{0.1truein}
\begin{tabular}{|c|c|c|c|c|c|}
\hline
Lattice & $D_s(0^-)$(MeV) & $D_s^*(1^-)$(MeV) & $D_{s0}^*(0^+)$(MeV)
& $D_{s1}(1^+)$(MeV) \\
\hline
$32^3\times 64$($m_{sea}a =0.006$) &1969&2121(9)&2304(22)&2546(27) \\
\hline
$16^3\times 72$ (quenched) &1976(11)&2131(13)&2248(78)&2476(92) \\
\hline
Experiment               &1968.49(34)&2112.3(5)&2317.8(6)&2535.35(84) \\
\hline
\end{tabular}
\end{center}
\end{table}

\begin{figure}[th]
\begin{center}
\caption{\bf Charmed Strange meson spectrum on $32^3\times 64$ 
$m_{sea}a=0.006$, DWF lattices  with overlap valence fermions}
\vspace{0.1truein}
\setlength{\unitlength}{0.240900pt}
\ifx\plotpoint\undefined\newsavebox{\plotpoint}\fi
\sbox{\plotpoint}{\rule[-0.200pt]{0.400pt}{0.400pt}}%
\begin{picture}(1349,809)(0,0)
\font\gnuplot=cmr10 at 10pt
\gnuplot
\sbox{\plotpoint}{\rule[-0.200pt]{0.400pt}{0.400pt}}%
\put(181.0,99.0){\rule[-0.200pt]{4.818pt}{0.400pt}}
\put(161,99){\makebox(0,0)[r]{1600}}
\put(1308.0,99.0){\rule[-0.200pt]{4.818pt}{0.400pt}}
\put(181.0,217.0){\rule[-0.200pt]{4.818pt}{0.400pt}}
\put(161,217){\makebox(0,0)[r]{1800}}
\put(1308.0,217.0){\rule[-0.200pt]{4.818pt}{0.400pt}}
\put(181.0,336.0){\rule[-0.200pt]{4.818pt}{0.400pt}}
\put(161,336){\makebox(0,0)[r]{2000}}
\put(1308.0,336.0){\rule[-0.200pt]{4.818pt}{0.400pt}}
\put(181.0,454.0){\rule[-0.200pt]{4.818pt}{0.400pt}}
\put(161,454){\makebox(0,0)[r]{2200}}
\put(1308.0,454.0){\rule[-0.200pt]{4.818pt}{0.400pt}}
\put(181.0,572.0){\rule[-0.200pt]{4.818pt}{0.400pt}}
\put(161,572){\makebox(0,0)[r]{2400}}
\put(1308.0,572.0){\rule[-0.200pt]{4.818pt}{0.400pt}}
\put(181.0,691.0){\rule[-0.200pt]{4.818pt}{0.400pt}}
\put(161,691){\makebox(0,0)[r]{2600}}
\put(1308.0,691.0){\rule[-0.200pt]{4.818pt}{0.400pt}}
\put(181.0,809.0){\rule[-0.200pt]{4.818pt}{0.400pt}}
\put(161,809){\makebox(0,0)[r]{2800}}
\put(1308.0,809.0){\rule[-0.200pt]{4.818pt}{0.400pt}}
\put(181.0,40.0){\rule[-0.200pt]{276.312pt}{0.400pt}}
\put(1328.0,40.0){\rule[-0.200pt]{0.400pt}{185.252pt}}
\put(181.0,809.0){\rule[-0.200pt]{276.312pt}{0.400pt}}
\put(40,424){\makebox(0,0){\rotatebox{90}{\bf Mass (MeV)}}}
\put(355,-18){\makebox(0,0){\small\bf $D_s(0^{-})$}}
\put(557,-18){\makebox(0,0){\small\bf $D_{s0}^*(0^+)$}}
\put(732,-18){\makebox(0,0){\small\bf $D_{s1}(1^+)$}}
\put(915,-18){\makebox(0,0){\small\bf $D_{s1}(1^+)$}}
\put(1099,-18){\makebox(0,0){\small\bf $D_s^*(1^-?)$}}
\put(319,377){\makebox(0,0){\small\bf $D_{s}(1969)$}}
\put(686,667){\makebox(0,0){\small\bf $D_{s1}(2460)$}}
\put(869,714){\makebox(0,0){\small\bf $D_{s1}(2536)$}}
\put(502,584){\makebox(0,0){\small\bf $D_{s0}^*(2317)$}}
\put(1053,461){\makebox(0,0){\small\bf $D_{s}^*(2112)$}}
\put(181.0,40.0){\rule[-0.200pt]{0.400pt}{185.252pt}}
\put(319,317){\usebox{\plotpoint}}
\put(319.0,317.0){\rule[-0.200pt]{21.922pt}{0.400pt}}
\put(502,524){\usebox{\plotpoint}}
\put(502.0,524.0){\rule[-0.200pt]{22.163pt}{0.400pt}}
\put(686,608){\usebox{\plotpoint}}
\put(686.0,608.0){\rule[-0.200pt]{21.922pt}{0.400pt}}
\put(869,652){\usebox{\plotpoint}}
\put(869.0,652.0){\rule[-0.200pt]{22.163pt}{0.400pt}}
\put(1053,399){\usebox{\plotpoint}}
\put(1053.0,399.0){\rule[-0.200pt]{21.922pt}{0.400pt}}
\put(410.0,317.0){\usebox{\plotpoint}}
\put(400.0,317.0){\rule[-0.200pt]{4.818pt}{0.400pt}}
\put(400.0,318.0){\rule[-0.200pt]{4.818pt}{0.400pt}}
\put(594.0,503.0){\rule[-0.200pt]{0.400pt}{6.263pt}}
\put(584.0,503.0){\rule[-0.200pt]{4.818pt}{0.400pt}}
\put(584.0,529.0){\rule[-0.200pt]{4.818pt}{0.400pt}}
\put(961.0,643.0){\rule[-0.200pt]{0.400pt}{7.709pt}}
\put(951.0,643.0){\rule[-0.200pt]{4.818pt}{0.400pt}}
\put(951.0,675.0){\rule[-0.200pt]{4.818pt}{0.400pt}}
\put(1144.0,402.0){\rule[-0.200pt]{0.400pt}{2.650pt}}
\put(1134.0,402.0){\rule[-0.200pt]{4.818pt}{0.400pt}}
\put(410,317){\circle*{18}}
\put(594,516){\circle*{18}}
\put(961,659){\circle*{18}}
\put(1144,407){\circle*{18}}
\put(1134.0,413.0){\rule[-0.200pt]{4.818pt}{0.400pt}}
\end{picture}
\end{center}
\end{figure}
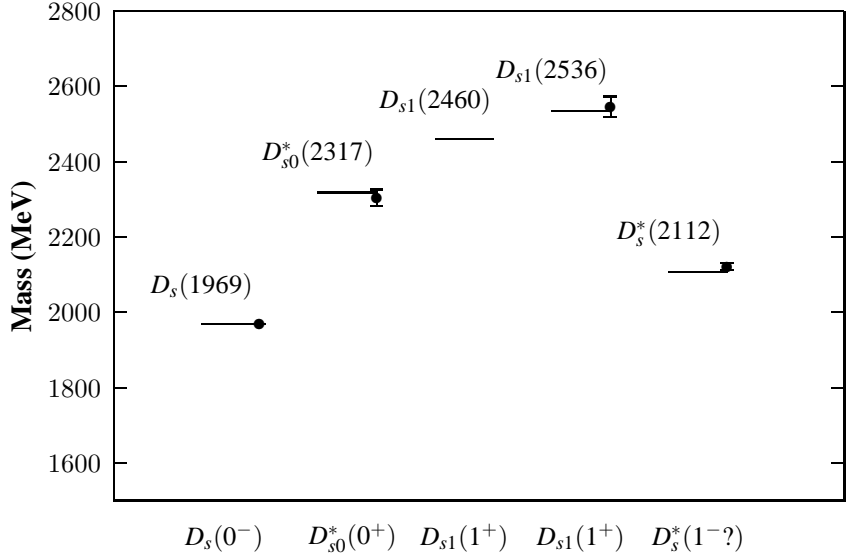

 This calculation is preliminary in that we have obtained results for one set
of sea quark masses and  one lattice spacing. We need to carry out calculation
with other sets of sea quark masses to do chiral extrapolation and on a coarser
$24^3 \times 64$ lattice in order to extrapolate to the continuum limit.  Nevertheless, the 
preliminary results show that the overlap fermion on  2+1 flavor domain wall
dynamical fermion configurations works reasonably well for the $D_S$ spectrum. The
results are consistent with experimental masses, especially for $D_{s0}^{~*}(2317)$, the
scalar meson. We don't see evidence that $D_{s0}^{~*}(2317)$ is an exotic meson. 
Rather, it fits well as a scalar $c\bar{s}$ meson.  

This work is partially supported by DOE Grant DE-FG05-84ER40154. J.B. Zhang
is partially supported by Chinese NSFC-Grants No.10675101 and 10835002.
We thank RBC and UKQCD collaborations for sharing the 2+1
flavor domain wall dynamical configurations with us.



\begin{thebibliography}{99}
\bibitem{BaBar}
B. Aubert et al., [BaBar Collaboration], Phys. Rev. Lett. 90, 242001 (2003); 
[arXiv: hep-ex/030402].
\bibitem{CLEO}
D. Besson, et al. [CLEO Collaboration], AIP Conf. Proc. 698:497497-502, 2004;
[arXiv: hep-ex/030517].
\bibitem{Godfrey}
S. Godfrey and I. Isgur, Phys. Rev. D32, 189 (1985); \\
S. Godfrey and R. Kokoski, Phys. Rev. D43, 1679 (1991).
\bibitem{bcl03} T. Barnes, F.E. Close, and H.J. Lipkin,
Phys. Rev. D68:054006 (2003), [arXiv: hep-ph/0305025].
\bibitem{Cheng}
H.Y. Cheng and W.S. Hou, Phys. Lett. B566, 193 (2003); [arXiv: hep-ph/0305038].
\bibitem{br03} 
E. van Beveren and G. Rupp, Phys. Rev. Lett., 91, 012003 (2003).
\bibitem{Lewis}
R. Lewis and R.M. Woloshyn, Phys. Rev. D62, 114507 (2000);
[arXiv: hep-lat/0003011].
\bibitem{Gunnar}
G.S. Bali, Phys. Rev. D68, 071501 (2003); [arXiv:hep-ph/0305209].
\bibitem{CP-PACS}
Y. Kayaba el al. [CP-PACS Collaboration], JHEP 0702, 019 (2007), [arXiv:
hep-lat/0611033].
\bibitem{amt06}
C. Allton, C. Maynard, A. Trivini, and R. Tweedie, PoS(LAT2006), 202 (2006), 
[arXiv:hep-lat/0610068]. 
\bibitem{Liu1}
K.F. Liu and S. J. Dong, Int. J. Mod. Phys.A20, 7241 (2005), [arXiv:
hep-lat/0206002].
\bibitem{Draper}
T. Draper et al. [$\chi$QCD Collaboration], [arXiv: hep-lat/0609034].
\bibitem{Liu2}
S.J. Dong and K.F. Liu, PoS(LAT2007), 093 (2007), [arXiv:0710.3038(hep-lat)].
\bibitem{Dong}
S.J. Dong and K.F. Liu, PoS(LAT2008), 117 (2008), [arXiv:0810.2993(hep-lat)].
\bibitem{JW}
J.W. Chen, D. O'Connell, and Andre Walker-Loud, Phys. Rev. D75, 054501 (2007);
C. Aubin, J. Laiho and R. Van de Water, Phys. Rev. D77, 114501 (2008),
[arXiv:0803.0129].
\bibitem{RBC} 
E.E. Scholz, RBC and UKQCD Collaborations, PoS(Lattice2008), 095 (2008), 
[arXiv:0809.3251v1(hep-lat)].
\bibitem{Sonali}
S. Tamhankar et al. [$\chi$QCD Collaboration], Phys. Lett. B638, 55 (2006),
[arXiv: hep-lat/0507027].

\end{thebibliography}
\end{document}